\RequirePackage{fix-cm}
\documentclass[twocolumn]{svjour3}        
\smartqed  
\usepackage{graphicx}
\usepackage{epsfig}
\usepackage{dcolumn}
\usepackage{bm}
\usepackage{color}
\usepackage{SIunits}
\usepackage{epstopdf}
\journalname{Experiments in Fluids}
\begin{document}

\title{Wide band Fresnel super-resolution \\applied to capillary break up of viscoelastic fluids}

\titlerunning{Wide band Fresnel super-resolution}

\author{Jorge E. Fiscina$^{1,2}$ \and Pierre Fromholz$^{1,3}$ \and Rainer Sattler$^{1}$ \\and Christian Wagner$^{1,*}$}

\institute{$^1$ Experimentalphysik, Universit\"at des Saarlandes, 66123 Saarbr\"ucken, Germany. \\$^2$  Gravitation Group, TATA institute of fundamental research, 1 Homi Bhabha Rd., 400005 Mumbai, India. \\$^3$ ICFP, Département de Physique, Ecole Normale Supérieure, 24, rue Lhomond 75005 Paris, France.
\\$^*$ email {c.wagner@mx.uni-saarland.de}}

\date{Received: date / Accepted: date}

\maketitle

\begin{abstract}
  We report a technique based on Fresnel diffraction with white illumination that permits the resolution of capillary surface patterns of less than $100$ nanometers. We investigate Rayleigh Plateaux like instability on a viscoelastic capillary bridge and show that we can overcome the resolution limit of optical microscopy. The viscoelastic filaments are approximately $20$ microns thick at the end of the thinning process when the instability sets in. The wavy distortions grow exponentially in time and the pattern is resolved by an image treatment that is based on an approximation of the measured rising flank of the first Fresnel peak.
\end{abstract}

\section{Introduction}
\label{intro}
Quantitative determination of the shape of capillary surfaces has been one of the standard problems in hydrodynamics, and several different techniques have been developed to deal with this. One typical example is the stationary pending droplet. From its surface profile one can deduce the surface tension \cite{Song1996}. Nowadays, digital imaging is used to determine the edges of the photographed object, and the intensity change at the surface is imaged with a limited amount of pixels on the camera chip. A standard high resolution procedure used to reconstruct the interface is based on an evaluation of the intensity gradient \cite{Montanero2008} by using a Gauss filter and a Sobel operator \cite{Scharr2000}. The latter is a discrete differentiation operator that calculates the gradient of the image intensity function. Typically, heuristic fitting functions, e.g. a sigmoid, are then used to model the interface, and the resolution can be improved by an educated guess on the model function. In this way, one goes already beyond the ideal view that the imaged contours correspond to a step function, and these procedures allow significant sub-pixel resolution
 \textcolor{black}{Based on a technique presented by Song and Spring  in $1996$ \cite{Song1996}, Vega, Montanero and Fernandez \cite{Vega2009}  proposed in $2009$  an image processing technique based on evaluating a local threshold and fitting the interface with a Boltzmann function. They investigated a vibrating hexahexane liquid bridge. They could detect oscillation amplitudes 20 times smaller than one pixel (80 nm for the objective they use: 1.6µm/pixel). During the same year Montanero, Vega and Ferrera reported a study about the micrometer deformation of a milli-metric liquid bridge, a liquid film of hexadecane deposited on a rod with an anti vibration isolation system \cite{Montanero2009}. They evaluated the standard deviation in the contour eccentricity to 50 nm and they attributed this the uncertainty associated to the image processing technique. The sub-pixel resolution reported here has a comparable resolution limit but it is fully based on a physical description of the light intensity field around the interface and thus allows to determine of the size of the object without a specific criterion to determine the threshold in the interface fitting function.}

\begin{figure}[htbp]
	\centering
		\includegraphics[width=0.9\linewidth]{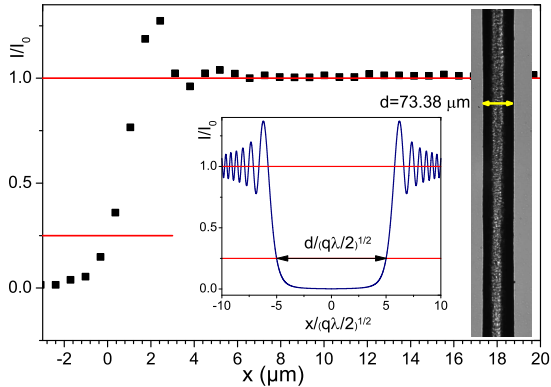}
	\caption{(color online) Right: gray level image of a human hair taken with a $20$ fold objective under white light illumination. Left: Measured diffracted light intensity pattern from a typical section of the right image. The straight red lines indicate the intensity level at infinity and at $1/4$ of this level. The one-fourth level indicates the actual position of the interface. The inset shows the diffraction pattern for the case of a fully absorbing object of diameter $d$, illuminated by monochromatic light of wavelength $\lambda$; $x$ is the distance perpendicular to the optical axis and q is related to the edge-detector distance (see text).}
	\label{Fresnelbeugung}
\end{figure}

There are different approaches to improving the optical resolution beyond Abbe’s diffraction limit \cite{Abbe1873}, often by evaluating the intensity profile of the diffracted light in the near field. A simple fitting function can be used to evaluate the shape of the interface. This super-resolution technique was being developed several years ago \cite{Pohl1984,Fischer1985,Durig1986,Lewis1984,Harrotunian1986} and applied successfully in near field scanning optical microscopy at a resolution down to 12 nm\cite{Betzig1992}.

In ref. \cite{Sattler2008,Sattler2010,Sattler2012} some of the authors of the present studies have shown that in extensional rheometry experiments one can resolve the amplitude of the sinusoidal deformation of the cylindrical liquid filament down to a resolution of $80$ nm. \textcolor{black}{This could be achieved by fitting the interface profile that was determined by a simple threshold algorithm with a sinus function, i.e. they had to assume a sinusoidal deformation of the interface.} A capillary break-up rheometer (CaBER) consists of two steel plates with a droplet of a viscoelastic sample liquid placed between them. The steel plates are drawn apart and due to the action of the capillary forces, the capillary bridge starts to shrink, and, at least in polymer solutions, one typically observes a long lasting filament that shrinks exponentially in time and eventually becomes unstable to a Rayleigh-Plateau like instability. The flow in the filament is purely elongational and the polymers are very efficiently stretched, therefore, the resistance to the flow is much stronger than in shear flow, which is described by the elongational viscosity. With knowledge of the surface tension one can deduce this elongational viscosity by simply evaluating the thickness of the filament as a function of time. In simple, Newtonian fluids, the elongational viscosity is simply three times the shear viscosity due to geometrical scaling; this is Trouton’s ratio. The aim of the present manuscript is to describe the optical procedure in more detail.

Our experimental method is closely related to what is called laser diffraction edge detection in industrial applications \cite{GmbH}. The edge position of an object from its Fresnel diffraction pattern can be detected either by illumination with a monochromatic light of wavelength $\lambda$ or with a polychromatic light of a given bandwidth $\Delta$$\lambda$. The diffraction pattern in the near field depends on the distance of the object from the source, and the full relation between the object and the electromagnetic field in the full space is quite complex. However, if one is interested only in the position of the edge, one can devise a simple formula from which information on the position of the object’s edge can be deduced, and where only the (averaged) distance between the imaged electromagnetic intensity distribution and the object has to be taken as a fitting parameter. The inset of Fig. \ref{Fresnelbeugung} shows the textbook case of Fresnel diffraction with monochromatic light at the edge of a fully absorbing object, as well as the distances of the intensity maxima and minima from the edge scale with 1/$\sqrt{\lambda q/2}$, where q is the distance between the object and (imaged) measuring position. It can be shown that the edge of the object is located at $1/4$ of its intensity at infinity $I_{0}$ \cite{Echt}. In Fig. \ref{Fresnelbeugung} the measured intensity distribution pattern around one edge is shown in the case of white light interacting with a human hair.

\section{Experimental Setup}
\label{Setup}
Our home-made CaBER consists of two steel discs with a diameter of $2mm$. The lower disc is fixed and the upper disc can be drawn apart with a linear motor (P01-23x80, Linmot, Spreitenbach, Switzerland). A $100W$ halogen lamp with a dichroic reflector is used to produce a shadow graph image of the capillary bridge. Optically, the fluid filament is similar to a cylindrical lens. The light beam that passes through the center of the filament is not diffracted and produces a bright line in the middle of the filament; the Light that passes close to the edges gets diffracted and so they appear dark. \textcolor{black}{The distance between the disks at maximal extension of the filament was $1.8$ mm.}

\begin{figure}[htbp]
	\centering
		\includegraphics[width=0.9\linewidth]{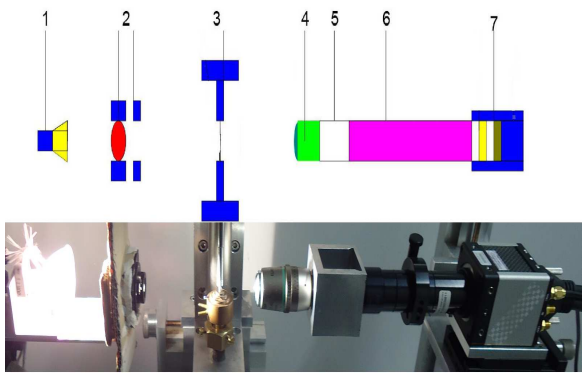}
	\caption{(Color online) Optical setup for the CaBER; 1: halogen lamp with dichroic reflector; 2: lens and iris; 3: two steel discs holding the liquid filament between them; 4: objective; 5: diaphragm/coupler; 6: collimator; 7: high speed camera.}
	\label{setup}
\end{figure}
The optical set-up for the CaBER is shown in Fig. \ref{setup}. The filament is filmed by a $10 bit$ high speed camera (X-Stream XS-5, IDT, Tallahassee, USA) at $1kHz$ frame rate and $1ms$ shutter time. The camera has $1280\times1024$ pixels with a size of $12\times12\mu m$. Microscope objectives (Nikon) with different magnifications of 10 and 20 fold were used. The latter was a long working distance objective ($WD$=7mm) with a numerical aperture of $NA$=0.45. At the shortest light wave length with reasonable sensitivity of the camera of $\Lambda\approx 450nm$ this yields a diffraction limited resolution of $\delta$=0.6$\mu$m. Such a length is magnified to 12$\mu$m on the CCD chip.

The depth of field was $5 \mu m$, which made it very difficult to obtain sharp images during the final stages. Filaments with a thickness of the order of $d \approx 10 \mu m$ near the end of the thinning process were very sensitive to the slightest distortions caused by air currents, or showed non reproducible lateral movements in the order of several microns even when air currents were mostly suppressed by an additional glass box around the capillary bridge.

Calibration with a reference object is needed in order to carry out this measurement method correctly. For this propose we use an optical glass fiber manufactured by nu-Fern, with specified diameter of $(130\mp1)\mu m$.

\begin{figure}[htbp]
\centering
\includegraphics[width=0.9\linewidth]{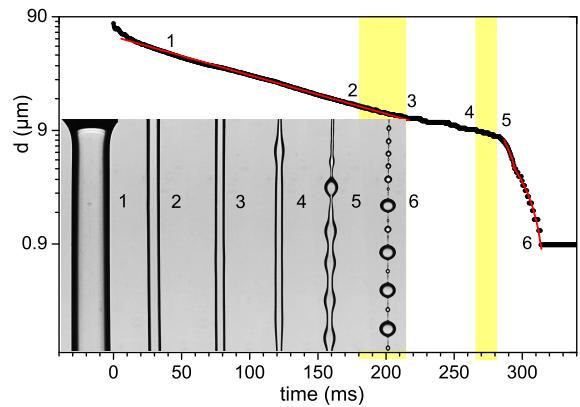}
\caption{(color online) Evolution of the neck diameter of the thinning capillary bridge of a droplet of 2000 ppm PEO solution in a CaBER experiment taken from reference \cite{Sattler2008}. The photographs correspond to the numbers indicated on the curve. In the interval 1 to 3 the curve is described by the exponential law $h_{min}(t)=h_{0}exp(-t/\tau)$ with $\tau= 130\pm30ms$ the characteristic time scale (red line). The events between 3 and 4 correspond to the formation of the first bead at the lower end of the filament. The first generation of beads on the filament appear between 4 and 5; from 5 to 6 the filament thins linearly.}
	\label{timeline}
\end{figure}

Our sample solution was a 1000 or 2000 ppm polyethylene oxide (PEO) with a Mw of $4\times 10^6$ amu dissolved in water. Its physical properties under elongation are reported in several publications \cite{Sattler2008,Sattler2012}. Figure \ref{timeline} shows the evolution of the fluid capillary bridge. After the filament is formed, one sees exponential shrinking until it becomes unstable to a Rayleigh-Plateau instability and then a final pinch off which is governed by self-similar laws. At this stage, the shrinking of the filament follows a linear law \cite{Eggers1993,EV08}. Both time periods, the exponential thinning between the numbers 1 and 2, and the linear thinning between 5 and 6 are indicated with the corresponding fitting curves (red lines), and the first wavy surface distortion on the filament (sinusoidal waves or single droplets) appears between the numbers 3 and 4; the first generation of beads is visible in the photographs from 4 and 5.

\section{Fresnel diffraction in the near field}
\label{Fresnel}
We analyze the intensity profile of diffracted light at the edge of an object, i.e. the capillary bridge. The intensity distribution of the electromagnetic light field
around the object is imaged by the microscope lens on the CMOS chip. \textcolor{black}{Our object is much larger than the wave length of the light} and we and we only have to evaluate the near field by use of Fresnel theory \cite{Echt} by considering a white illumination.  The calculation should take into account the integration over the spectral range that is determined by the CMOS chip sensitivity. In our case, it is the range from $300$ to $800$ nm, with a peak around $550$ nm and still non negligible sensitivity in the near infra-red.

In principle, for the light source we have to consider the differential radiant energy density of a black body

\begin{equation}\label{diffstrdichte}
S(\nu)=2h\nu^{3}c^{-2}\left (e^{\frac{h\nu}{k_{B}T}}-1\right)^{-1},
\end{equation}

with $h$ the Planck's constant, $k_B$ the Boltzmann's constant, $c$ the speed of light, $\nu$ the frequency of the light, and we estimate a temperature of $T$=3000K. However, our halogen spot is equipped with a dichroic mirror and in this combination the spectrum is better approximated by the rectangular window function, which is $1$ around a central wave length $\lambda$ within the wide band $\Delta$$\lambda$ and $0$ outside this range.

We define the two Fresnel integrals

\begin{eqnarray}\label{FresnelIntegrale}
 C(x)=\int_{0}^{x}cos(\frac{\pi t^{2}}{2})dt\\
 S(x)=\int_{0}^{x}sin(\frac{\pi t^{2}}{2})dt.
\end{eqnarray}

When the midsection of the liquid filament is at a distance p from a
point source, the intensity distribution at a point P(q,x) with q
being the distance from the filament along the optical axes and $x$ the one
perpendicular to it is given by
\begin{equation}\label{InF}
    I=I_0 \frac{1}{2}(\frac{1}{2}-C(X))^{2}+\frac{1}{2}(\frac{1}{2}-S(X))^{2},
\end{equation}

with $I_0$ the intensity of the incoming light and the auxiliary function

\begin{equation}\label{nu2f}
X(\lambda,p,q,x)=\frac{-\sqrt{2(p+q)}x}{\sqrt{\lambda pq}}.
\end{equation}

For our case where the light source is far away from the filament compared to the near field ($q<<<p$) it follows that:

\begin{equation}\label{nu2f}
X(\lambda,p,x)=\frac{-x}{\sqrt{\lambda q/2}}.
\end{equation}

For the experimental data evaluation q is taken as a fit parameter. In principle one would have to integrate q over the thickness of the depth of field but our main interest lies in the steep first flank of the first Fresnel peak that is not sensitive to this integration procedure and instead we can fit a kind of effective q.

\begin{figure}[htbp]
	\centering
		\includegraphics[width=0.9\linewidth]{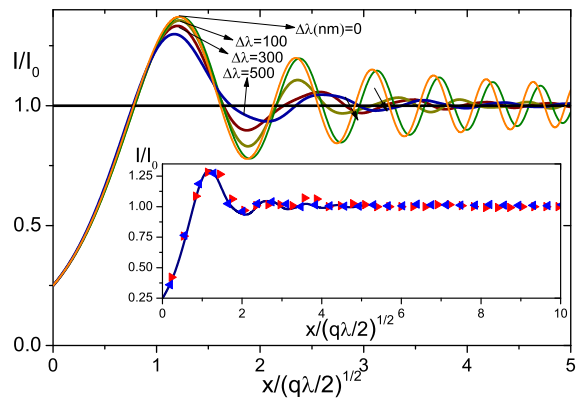}
	\caption{(color online) Intensity distribution resulting from the interaction of a light field and the edge of a fully absorbing object for different band with of the light source $\Delta \lambda$ . The inset shows the measurement for a section of the human hair of Fig. \ref{Fresnelbeugung} by folding the two edges together (right and left triangles) and comparing with the calculated pattern for a band width $\Delta$$\lambda$= $500$ nm around the peak intensity wave length of $550nm$ of our CMOS-chip.}
	\label{WideBand}
\end{figure}

\begin{figure}[htbp]
	\centering
		\includegraphics[width=0.9\linewidth]{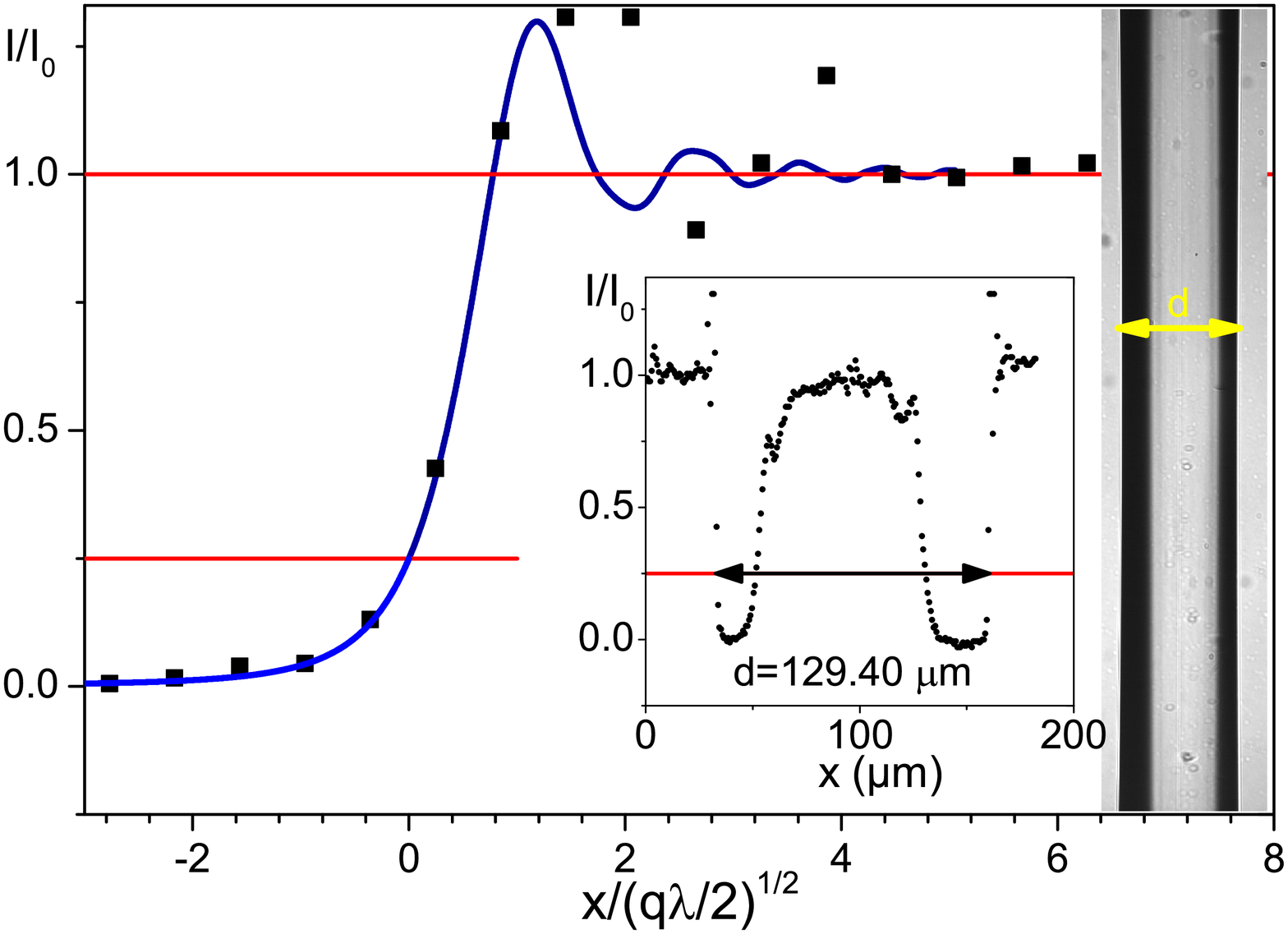}
	\caption{(color online) Right: gray level image of an optical fiber taken with a $20$ fold objective under white light illumination. Left: Measured points for one of the edges of the optical fiber and calculated (line) intensity distribution around the optical fiber for a band width $\Delta$$\lambda=500$ nm with central wave length $\lambda=550$ nm. The inset shows the full width of the fiber from which the thickness can be determined.}
	\label{Fresnelfiber}
\end{figure}

We still have to determine the bandwidth $\Delta \lambda$ of the light source; the set of curves for different values of $\Delta \lambda$ is shown in Fig. \ref{WideBand}. Again we find that the rising flank of the first Fresnel peak is not sensitive to $\Delta \lambda$. In the inset of Fig. \ref{WideBand} the theoretical curve for $\Delta \lambda = 500 nm$ is compared to the experimental result for the profile of a human hair from Fig. \ref{Fresnelbeugung}. Similar agreement between the measured points and these calculations is observed for an optical fiber in Fig. \ref{Fresnelfiber}.

\section{Calibration with a reference object}
\label{Calibration}

As q is a fitting parameter, the technique works best if we calibrate the set-up with an object of known dimensions and geometry. Therefore, we measure an optical fiber (Nu-Fern, with well known specifications (Fig. \ref{Fresnelfiber}). The calibration is not only to introduce the length scale into our algorithm, but also for mapping the optical aberration of the $20$ fold objective.

\begin{figure}[h]
	\centering
		\includegraphics[width=0.9\linewidth]{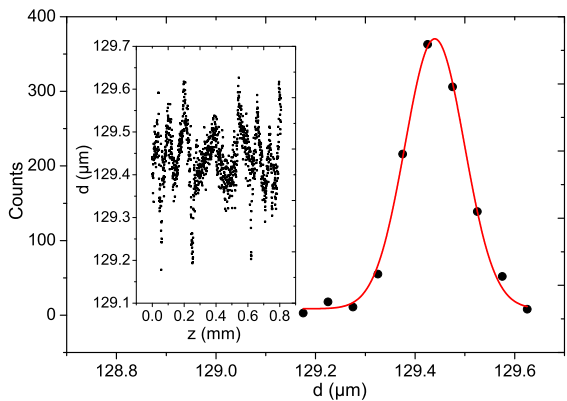}
	\caption{(color online) Gaussian distribution of frequency analysis of the optical fiber diameter profile. The inset shows this profile; the diameter $d$ along the $z$- axis of the fiber shown in Fig. \ref{Fresnelfiber}.}
	\label{Glassfiber}
\end{figure}

For this calibration, in front of the objective we set up a hanging optical fiber with a weight glued on one end that acts as a plumb line. This allow us to assume the fiber is a straight line, thus we are able to map the optical aberration by moving this vertical in front of the lens. Once we get this map of the aberration, we use it to correct the image. The inset of Fig. \ref{Glassfiber} shows the corrected profile of the optical fiber after the identification and mapping of the objective pincushion aberration. Around the average diameter of the fiber of $129.4\mu m$,\textcolor{black}{ we observe an oscillatory distribution of intensity which we attribute to a higher order aberration of our optical system}. The distribution calculated from this profile of the fiber can be fitted with a Gaussian (Fig. \ref{Glassfiber}), where the main peak corresponds to the diameter $129.4\mu m$, with a $\sigma$=$59.8$ nm. The measured value is in good agreement with the manufacturer’s specifications $(130\mp1)\mu m$.  \textcolor{black}{The width of this distortion sets the limit for the resolution of our set-up.}

\begin{figure}[h]
	\centering
		\includegraphics[width=0.9\linewidth]{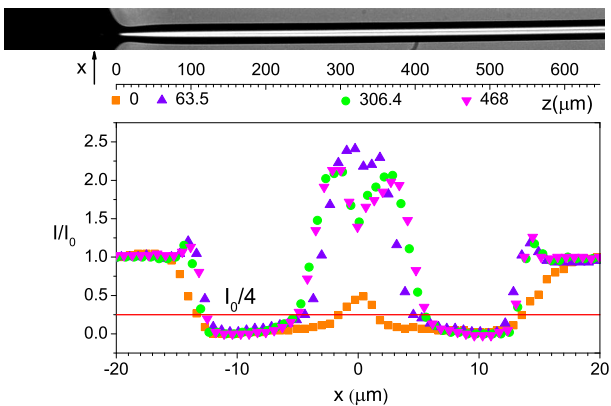}
	\caption{(color online) Top: gray level image of a liquid filament and the corresponding length scale along it; taken with a 20 fold objective during the thinning capillary bridge of a droplet of $1000$ ppm PEO. Bottom: Measublack intensity distribution across filament sections ($x$-axis) for different distances along the filament ($z$-axis) as they are indicated in the top image at $z=0 \mu m$ (square), $z=63.5 \mu m$ (up-triangle), $z=306.4 \mu m$ (circle) and $z=468 \mu m$ (down-triangle). The straight line at $1/4$ of the intensity at infinity indicates the level used for the super-resolution algorithm to measure the thickness of the filament.}
	\label{Fresnelprofile}
\end{figure}

\begin{figure}[h]
	\centering
		\includegraphics[width=0.9\linewidth]{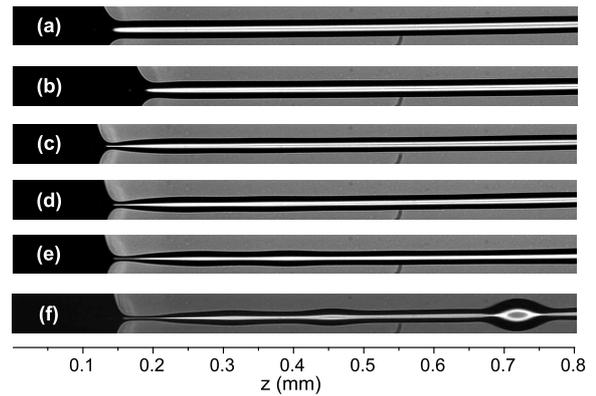}
	\caption{Gray level images of the liquid filament taken with a $20$ fold objective under white light illumination. They show a sequence in the time period of the wavy pattern that gives origin to the first generation of beads on the liquid filament. By taking the onset of the wavy distortion at t=0 ms in (a), they correspond to the times (b) $t=2.33$ ms, (c) $t=14.33$ ms, (d) $t=15$ ms, (e) $t=17.67$ ms, (f) $t=20.83$ ms.}
	\label{sequence}
\end{figure}

\begin{figure}[h]
	\centering
		\includegraphics[width=0.8\linewidth]{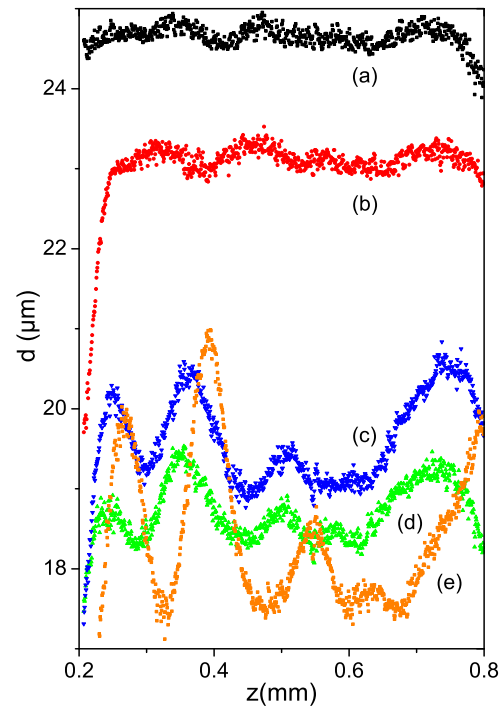}
	\caption{(color online) Sequence of curves obtained by applying the super-resolution algorithm to the temporal sequence of photographs of Fig. \ref{sequence} (a to e).}
	\label{Fresnelsequence}
\end{figure}

\section{Application to extensional rheology}
\label{Appl}

In this section we report the performance of our super-resolution technique by measuring the thinning capillary bridge of a droplet of $1000$ ppm PEO. Of particular interest is the time period involving the formation of the first droplets or beads on the filament; this period was indicated in Fig. \ref{timeline} by the range from 3 to 4. Fig. \ref{Fresnelprofile} shows a typical example of the intensity diffraction patterns for different sections across the mentioned liquid bridge or filament. The physical mechanism of the capillary break-up of this solution is described in detail in ref. \cite{Sattler2008,Sattler2012} while here we concentrate on the application of this edge-detection technique. Our algorithm evaluates the Fresnel diffraction pattern in the near field as described above. From the intensity plateaus on both sides of the border it is possible to identify the intensity level at infinity $I_{0}$. The edges of the filament are situated at the intensity level $I/I_{0}=1/4$.

The time period of the formation of the first droplets is shown in the sequence of Fig. \ref{sequence}. A wavy distortion is the beginning of the formation of beads on the filament. In the first photographs in Fig. \ref{sequence} our algorithm detects a wavy pattern with an amplitude of approximately 100nm in this image (Fig. \ref{Fresnelsequence} (a)). This  pattern grows in time and moves along the filament. In Fig. \ref{sequence}(e) it begins to be visible when looking at the photo with the naked eye, and it is very clear in Fig. \ref{sequence}(f). The correspondence between the peaks for each profile reveals that the beads are the wavy pattern as observed in Fig. \ref{Fresnelsequence} (a) to (e).

\begin{figure}[h]
	\centering
		\includegraphics[width=0.9\linewidth]{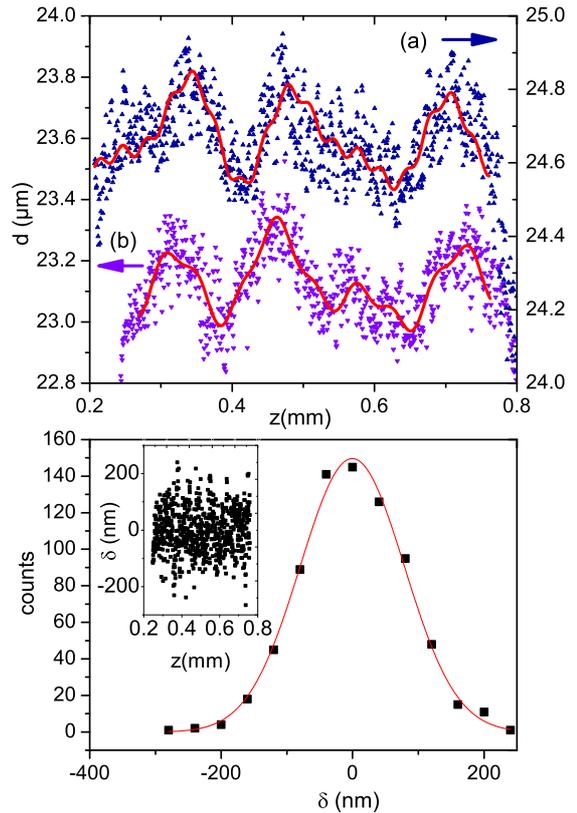}
	\caption{(color online) Top: Diameter profiles of the photographs of Fig. \ref{sequence} (a) and (b) and corresponding fittings with equation \ref{SumSin}. Bottom: Gaussian distribution that results from the diameter $d$ profiles.}
	\label{hundertnanometers}
\end{figure}

To evaluate the resolution of the super-resolution algorithm, we fit the profiles of the photographs (a) and (b) of the sequence of the Fig. \ref{sequence} with a sum of sinus, Eq. (\ref{SumSin}) (see Fig. \ref{hundertnanometers} top).

\begin{equation}\label{SumSin}
d=d_{0}+\sum_{i=1}^{5}A_{i}.\sin \left( \frac{\pi .\left( z_{i}-c_{i}\right)
}{w_{i}}\right)
\end{equation}

We perform the frequency analysis after subtracting this fitting function from the data and we obtain the corresponding distribution for the noise (see Fig. \ref{hundertnanometers} bottom). This results in a Gaussian distribution with a $\sigma$=$93$ nm. In spite of the noise, the center of the mass of the wavy oscillation can be resolved and described, for example, with amplitudes $A_{i}$ no larger than $100$ nm and fitting errors of $(\mp5)$ nm.

This method was used by Sattler et.al. in ref \cite{Sattler2008} without a detailed description of the optical method.

\section{Conclusions}
\label{Conclusions}
We have described a new experimental  method to evaluate the surface profile of a capillary interface with high precision. The method is based on the evaluation of the wide band Fresnel diffraction in the near field. The set-up has no specific requirements provided that good microscope objectives with a reasonable numerical aperture are used. We have used this method to characterize two model objects, a human hair and a glass fiber and we have shown a hydrodynamic application on a capillary extensional rheometer. With our technique it is possible to reach sub-pixel resolution for spatial variances with less than $100$ nm.

\textcolor{black}{The calibration with the optical fiber shows that the limit for the resolution is related to higher order optical aberration of our optical system that are of the order of $60$ nm. We assume that by improving the illumination conditions and the quality of the lens that it should be possible to reach even better resolution.}

\begin{acknowledgements}
This work was supported by the DFG-Project WA 1336 and Thermo Haake. JF thanks the Alexander von Humboldt foundation, Global Site S.L., and Mrs. Audrey Shaw for the English revision.
\end{acknowledgements}

\end{document}